\documentclass[paper]{JHEP3}
\usepackage{epsfig}
\usepackage{bbm}
\usepackage{amsmath}
\usepackage{graphicx}
\usepackage{cite}
\def\sla#1{\ifmmode%
\setbox0=\hbox{$#1$}%
\setbox1=\hbox to\wd0{\hss$/$\hss}\else%
\setbox0=\hbox{#1}%
\setbox1=\hbox to\wd0{\hss/\hss}\fi%
#1\hskip-\wd0\box1 }
\newcommand{\be}{\begin{equation}}
\newcommand{\ee}{\end{equation}}
\newcommand{\bea}{\begin{eqnarray}}
\newcommand{\eea}{\end{eqnarray}}

\title{\begin{boldmath}Gluon-induced $W$-boson pair production at the
    LHC
\end{boldmath}}

\author{T.~Binoth\\
School of Physics, The University of Edinburgh, Edinburgh EH9 3JZ, 
Scotland}

\author{M.~Ciccolini\\
Paul Scherrer Institut, CH-5232 Villigen PSI, Switzerland}

\author{N.~Kauer\\
Institut f\"ur Theoretische Physik, Universit\"{a}t W\"{u}rzburg,
D-97074 W\"{u}rzburg, Germany}

\author{M.~Kr\"{a}mer\\
Institut f\"{u}r Theoretische Physik E, RWTH Aachen, D-52056
Aachen, Germany}

\abstract{Pair production of $W$ bosons constitutes an important
  background to Higgs boson and new physics searches at the Large
  Hadron Collider LHC. We have calculated the loop-induced
  gluon-fusion process $gg \to W^{*}W^{*} \to {\rm leptons}$,
  including intermediate light and heavy quarks and allowing for
  arbitrary invariant masses of the $W$ bosons. While formally of
  next-to-next-to-leading order, the $gg \to W^{*}W^{*} \to {\rm
    leptons}$ process is enhanced by the large gluon flux at the LHC
  and by experimental Higgs search cuts, and increases the
  next-to-leading order $WW$ background estimate for Higgs searches by
  about $30\%$.  We have extended our previous calculation to include
  the contribution from the intermediate top-bottom massive quark loop
  and the Higgs signal process. We provide updated results for cross
  sections and differential distributions and study the interference
  between the different gluon scattering contributions. We describe
  important analytical and numerical aspects of our calculation and
  present the public {\tt GG2WW} event generator.}


\preprint{Edinburgh 2006/23\\
          PITHA 06/12\\
          PSI-PR-06-13}

\begin{document}


\section{Introduction\label{intro-section}}
Vector-boson pair production provides an important background to Higgs
boson searches in the $pp \to H \to W^{\ast}W^{\ast} \to {\rm
  leptons}$ channel at the Large Hadron Collider (LHC).  Since for
dileptonic $W$ decays no Higgs mass peak can be reconstructed, this
background cannot be estimated from measured data via sideband
interpolation.  Precise theoretical predictions for the irreducible
$W$-pair continuum background are hence crucial.

The hadronic production of $W$ pairs has been investigated extensively
in the literature (see e.g.\ Ref.~\cite{Haywood:1999qg}). The
next-to-leading order (NLO) QCD corrections to $q\bar{q} \to WW \to
\ell\bar{\nu}\bar{\ell'}\nu'$ have been known for some
time~\cite{Ohnemus:1991kk, Frixione:1993yp, Ohnemus:1994ff,
  Dixon:1998py, Dixon:1999di}.  More recently also single-resonant
contributions have been included \cite{Campbell:1999ah}, and the NLO
calculation has been matched with a parton shower
\cite{Frixione:2006he} and combined with a summation of soft-gluon
effects~\cite{Grazzini:2005vw}.  Electroweak corrections, which become
important at large $WW$ invariant masses, have been computed in
Ref.~\cite{Accomando:2004de}.

In this article we present the first complete calculation of the
gluon-induced process $gg \to W^{*}W^{*} \to {\rm leptons}$ and study
its importance as a background to Higgs searches in the $pp \to H \to
WW \to \ell^+\ell^-\sla{p}_T$ channel.  The gluon-induced background
process is mediated by quark loops and thus suppressed by two powers
of $\alpha_{\rm s}$ relative to quark-antiquark annihilation.
Although it formally enters only at next-to-next-to-leading order, the
importance of the gluon-fusion process is enhanced by experimental
Higgs search cuts.  These cuts exploit the longitudinal boost and the
spin correlations of the $WW$ system to suppress $W$-pair continuum
production through quark-antiquark
annihilation~\cite{Dittmar:1996ss,Dittmar:1996sp}.

The gluon-fusion contribution to on-shell $W$-pair production, $gg \to
WW$, has been computed in
Refs.~\cite{Glover:1988fe,Kao:1990tt,Duhrssen:2005bz}.  Here, we
present a fully differential calculation of gluon-induced $W$-boson
pair production and decay, $gg \to W^{\ast}W^{\ast} \to
\ell\bar{\nu}\bar{\ell'}\nu'$, including the top-bottom massive quark
loop contribution and the intermediate Higgs contribution with full
spin and decay angle correlations and allowing for arbitrary invariant
masses of the $W$ bosons.\footnote{Gluon-induced tree-level processes
  of the type $gg \to WW q\bar q$ are expected to be strongly
  suppressed~\cite{Adamson:2002jb,Adamson:2002rm} and have thus not
  been taken into account.  We note that results for a similar
  process, $gg \to Z^{\ast}Z^{\ast} \to 4 l^{\pm}$, have been
  presented in Refs.~\cite{Matsuura:1991pj,Zecher:1994kb} including
  massive quark contributions, correlated decays and off-shell
  effects.}  Partial results of this work have already been presented
in Refs.~\cite{Binoth:2005ua,BCKK_LH}.  In Ref.~\cite{Binoth:2005ua}
we found that the contributions of the first and second quark
generations enhance the NLO $WW$ background prediction for Higgs
searches by approximately $30\%$.

In the following we describe details of our calculation, introduce the
{\tt GG2WW} program and present cross sections and differential
distributions.  We discuss the impact of the third-generation
contribution and interference effects between massless and massive
quark loop as well as signal and complete background contributions.


\section{Calculation \label{calc-section}}

\subsection{Amplitude calculation preliminaries\label{prelimi}}
The calculation of the 1-loop amplitude for $gg
\to \ell\bar{\nu}\bar{\ell'}\nu'$ is sufficiently complex that it is advantageous
to organize Feynman amplitudes using form factors of tensor integrals, 
which are then evaluated numerically. This approach works well as long as 
the numerical representation of the amplitude is stable. When calculating
cross sections for tree-level processes at NLO, 1-loop amplitudes are interfered with
tree-level amplitudes. The cross section for our loop-induced process, however, is 
proportional to a squared 1-loop amplitude. Standard loop amplitude representations
will thus lead to more severe numerical instabilities.
It is therefore advantageous to employ an algebraic approach to
tensor reduction that maximizes the number of cancellations that occur at the
analytical level.  To control the size of the analytical expressions it
is necessary to split the amplitude into irreducible building blocks.
Thus gauge cancellations and compensations of unphysical
denominators in subexpressions of the full amplitude are facilitated, and one can use 
standard algebraic programs like {\sc Maple} and {\sc
  Mathematica} to factorize and simplify the expressions.
The calculation of the amplitude proceeds in the following steps:
\begin{itemize}
\item[--] translation of Feynman diagrams to amplitude expressions; 
\item[--] amplitude organization;
\item[--] evaluation of amplitude expressions and algebraic reduction;
\item[--] simplification of irreducible amplitude coefficients;
\item[--] numerical amplitude evaluation.
\end{itemize}
Before describing those steps in more detail we set up our notation
and provide some basic definitions.

We consider gluon-induced $W$-pair production and decay and thus calculate the 
parton amplitude 
\bea g(p_1,\lambda_1) +
g(p_2,\lambda_2) + \ell(p_5,-) + \bar\nu(p_6,+) +
\nu'(p_7,-) + \bar{\ell'}(p_8,+) \to 0\,,  \nonumber\eea 
where $\ell$ and $\ell'$ are charged, approximately massless leptons of different flavour and 
all momenta are ingoing.  This amplitude is related to the physical amplitude
by crossing symmetry.
$\lambda_{1,2}$ specify the gluon helicities and $p_{3,4}^2\equiv s_{3,4}$ the virtualities
of the off-shell vector bosons.  The coupling of the gluons to the vector
bosons is mediated through a quark loop.
Although six external particles are involved in the process, at most 1-loop
4-point functions occur in the calculation, because a pure QCD initial state couples 
to a pure electroweak final state.  The contributing
topologies are presented in Fig.~\ref{graphs}.
\unitlength=1mm
\FIGURE{\label{graphs}
\begin{picture}(150,75)
\put(5,65){a)}
\put( 15,38){\epsfig{file=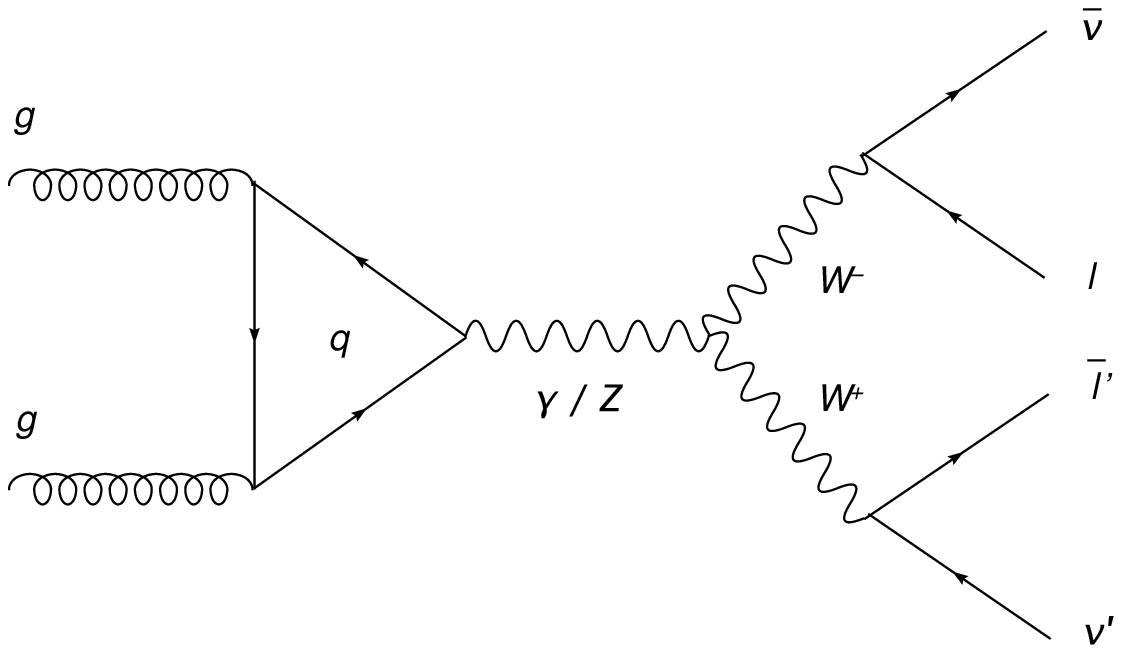,height=3.5cm}}
\put(80,65){b)}
\put( 90,38){\epsfig{file=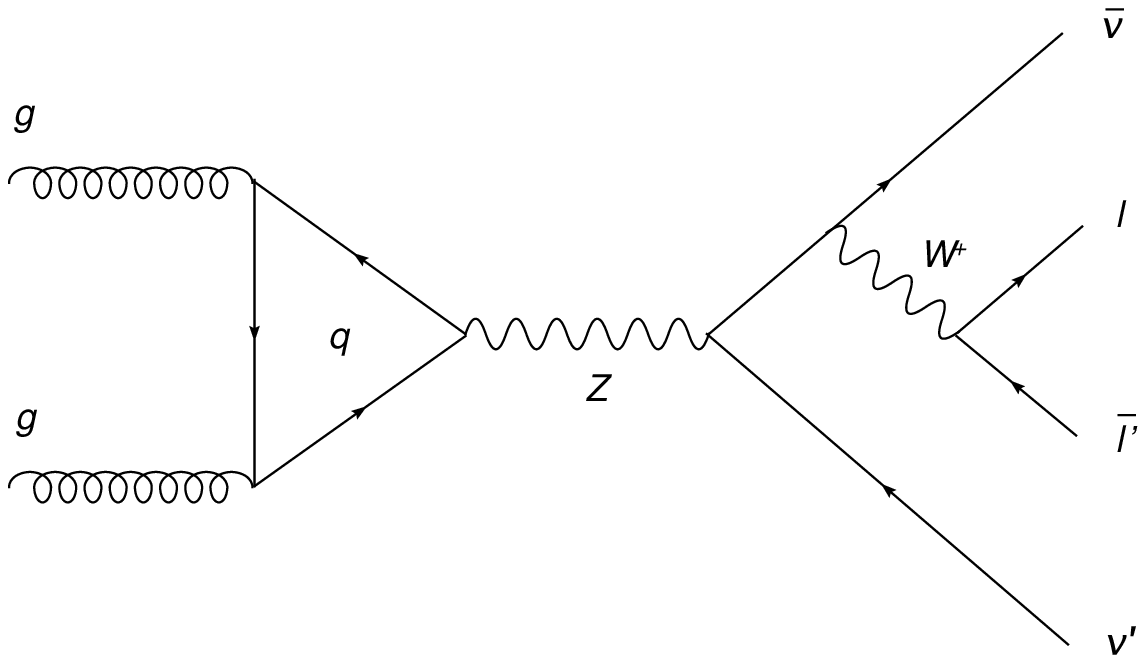,height=3.5cm}}
\put(5,30){c)}
\put( 15,0 ){\epsfig{file=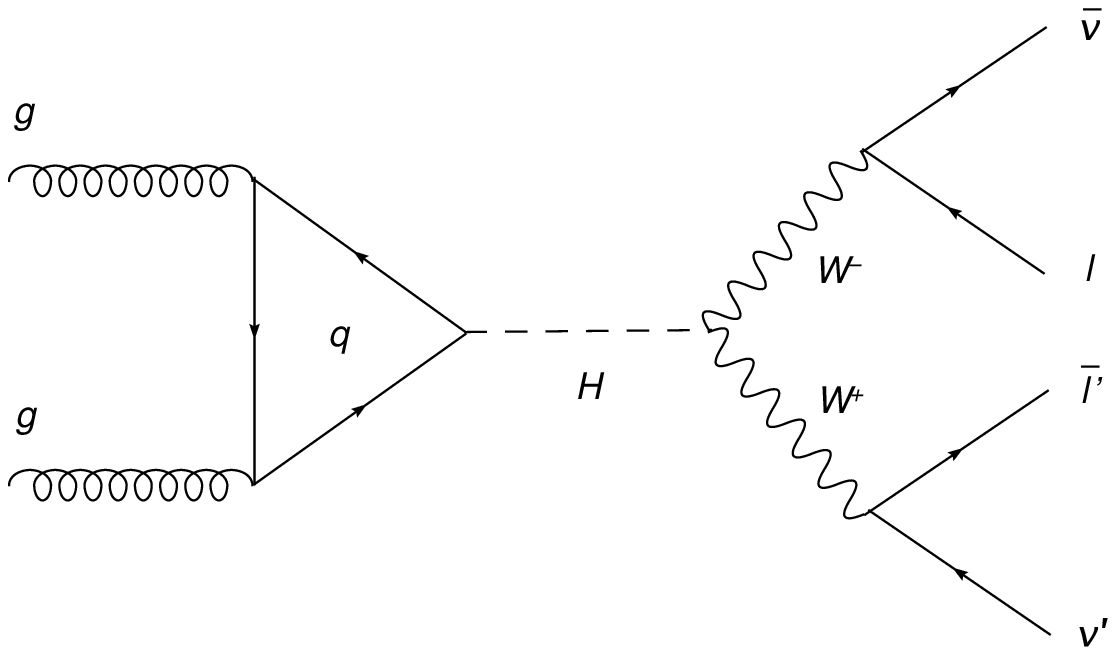,height=3.5cm}}
\put(80,30){d)}
\put( 90,0 ){\epsfig{file=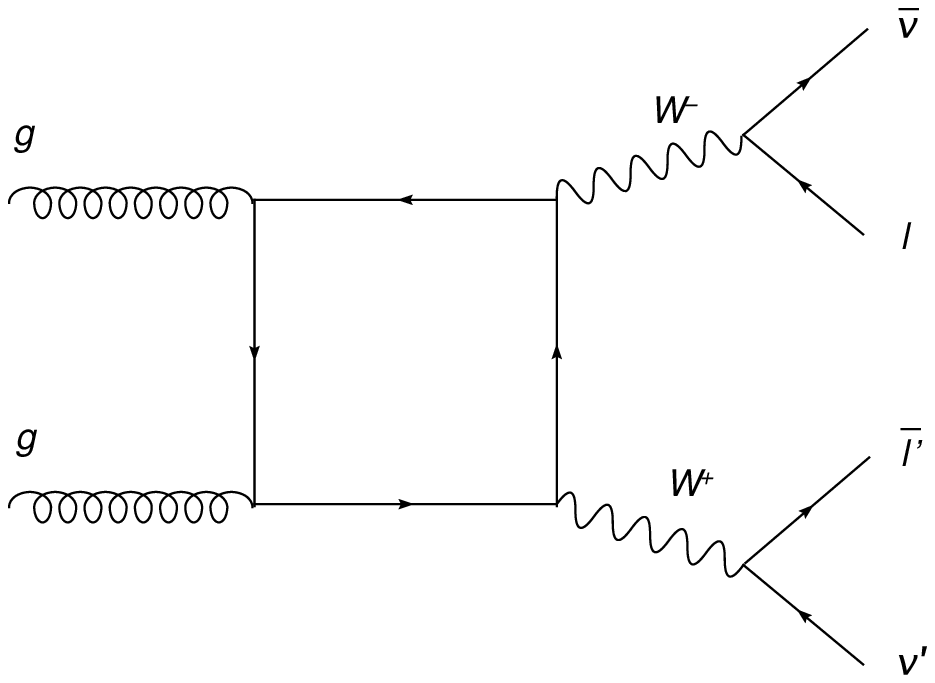,height=3.5cm}}
\end{picture}   
\caption{Examples for topologies contributing to $gg
\to \ell\bar{\nu}\bar{\ell'}\nu'$.
The $Z$-exchange 
  triangle diagrams cancel when summed.  Note that all external momenta are incoming.}}

We neglect masses for the first two quark generations and all leptons, and set the CKM
matrix to unity. The photon exchange graphs vanish due to Furry's theorem.
The $Z$-exchange diagrams, however, contain an axial coupling and are proportional to
$(m_u^2-m_d^2)(s_3-s_4)$, when summed over up- and down-type contributions.\footnote{Note that including the
single-resonant diagrams in Fig.~\protect\ref{graphs}b) is essential to
maintain gauge invariance.}
We find that they also vanish for massless quarks 
as required by Furry's theorem and weak isospin invariance.
If $m_u\neq m_d$, this argument is no longer valid and the 
triangle graphs could contribute.  
Note that in the on-shell case one has $s_3=s_4=M_W^2$, and they still vanish~\cite{Kao:1990tt}.  For arbitrary invariant masses $s_3\neq s_4$, we find that the contributions from double- (Fig.~\ref{graphs}a) and single-resonant (Fig.~\ref{graphs}b) diagrams with internal $Z$ propagator cancel each other if the decay leptons are massless (as assumed).  
The only triangle graphs that contribute are thus the Higgs exchange
diagrams (Fig.~\ref{graphs}c) 
with amplitudes proportional to the $qqH$ Yukawa couplings.
The box diagrams do not involve these couplings
and therefore form a gauge invariant subset. 
Since only double-resonant diagrams contribute, 
the amplitude factorizes into $W^\ast W^\ast$ production and 
subsequent decay mediated by the chiral fermion currents 
\bea 
J_3^{\mu} &=& \frac{1}{2} \, \bar v(p_6) \gamma^{\mu}
(1-\gamma_5) u(p_5)\ , \nonumber\\
J_4^{\mu} &=& \frac{1}{2} \, \bar v(p_8) \gamma^{\mu}
(1-\gamma_5) u(p_7)\ . \nonumber
\eea 
We note that the gauge-parameter dependent terms of the amplitude in $R_\xi$ gauge 
vanish for massless leptons due to current conservation.
The $W$ propagators thus simplify to Feynman-gauge form
\bea
P^{\mu\nu}(p) = \frac{-i\, g^{\mu\nu}}{p^2 - M_W^2 + i\,M_W
  \Gamma_W }\,, \nonumber
\eea   
since the Goldstone bosons do not couple to the massless external leptons.

We calculate the contributing helicity amplitudes using the spinor formalism of 
Ref.~\cite{xuetal}. The complexity of the calculation is governed by
the number of independent scales that occur, i.e. six in the case at hand.
We choose the Mandelstam variables $s=(p_1+p_2)^2$, $t=(p_2+p_3)^2$ and
$u=(p_1+p_3)^2$, and the $W$ virtualities, $s_3$ and $s_4$, which obey the relation 
$s+t+u=s_3+s_4$.  To account for the third generation, 
we calculate with non-zero quark masses $m_t$ and $m_b$.\footnote{
While keeping the full $m_b$ dependence in our calculation, 
we note that the limit $m_b \to 0$ is a very good approximation
for LHC energies.  The induced error is ${\cal O}(m_b^2/m_t^2) \sim 0.1\%$.  
As $m_b$ serves as an IR cutoff many basis function coefficients vanish
in this limit. The non-zero coefficients simplify considerably, too.}

Following Ref.~\cite{xuetal}, we use $p_2$ ($p_1$) as reference vector
for the polarization vector $\varepsilon_1$ ($\varepsilon_2$) and write
\bea
&& \varepsilon^+_{\mu}(p_1) = \frac{1}{\sqrt{2}} 
\frac{\langle  2^-|\mu|1^-\rangle }{\langle  2^-|1^+\rangle } \:,\:
\varepsilon^+_{\mu}(p_2) = \frac{1}{\sqrt{2}} 
\frac{\langle  1^-|\mu|2^-\rangle }{\langle  1^-|2^+\rangle }\ , \nonumber \\
&&\varepsilon^-_{\mu}(p_1) = \frac{1}{\sqrt{2}} 
\frac{\langle  2^+|\mu|1^+\rangle }{\langle  1^+|2^-\rangle } \:,\:
\varepsilon^-_{\mu}(p_2) = \frac{1}{\sqrt{2}} 
\frac{\langle  1^+|\mu|2^+\rangle }{\langle  2^+|1^-\rangle }\ , \nonumber \\
&& J_3^\mu =  \langle 6^-|\mu|5^-\rangle  \quad , \quad 
J_4^\mu =  \langle 8^-|\mu|7^-\rangle \, , \nonumber
\eea
and obtain as projectors for the $++$ and $+-$ helicity combinations:
\bea
\varepsilon_{1}^{+\;\mu} \varepsilon_{2}^{+\;\nu}&=& 
- \frac{[21]}{\langle 12\rangle } \frac{1}{s} 
\Bigl( 
p_1^\mu p_2^\nu + p_1^\nu p_2^\mu - p_1\cdot p_2 \; g^{\mu\nu} - i \, 
\epsilon^{\sigma\nu\rho\mu}p_{1\sigma}p_{2\rho}
\Bigr) \ ,\label{helproj} \\ 
 \varepsilon_1^{+\;\mu} \varepsilon_2^{-\;\nu} &=& 
-\frac{ \langle 23\rangle [31] }{\langle 13\rangle [32] }\;
 \frac{  \rm{tr}^-[ \sla{p}_1\sla{p}_3\sla{p}_2\gamma^\mu ]\  
\rm{tr}^-[\sla{p}_1\sla{p}_3\sla{p}_2\gamma^\nu ] }{2\,s (u\, t - s_3\,s_4)}\, ,
\label{gram_det_danger}
\eea
with $\textrm{tr}^-[\Gamma]\equiv ( \textrm{tr}[\Gamma] - \textrm{tr}[
\gamma_5 \Gamma ])/2$ and the spinor inner products $\langle ij\rangle\equiv\langle p_i^-|p_j^+\rangle,\ [ij]\equiv\langle p_i^+|p_j^-\rangle$, where $|p_i^\pm\rangle$ is the Weyl spinor for a massless particle with momentum $p_i$.  
We define the epsilon tensor by 
$4i\, \epsilon^{\mu\nu\rho\sigma} = 
{\rm tr}[\gamma_5\gamma^\mu\gamma^\nu\gamma^\rho\gamma^\sigma]$.
The spinor prefactors in Eqs.~(\ref{helproj}) and (\ref{gram_det_danger}) are pure phases and can be disregarded 
when calculating $|{\cal M}|^2$.

\subsection{Symbolic amplitude evaluation with algebraic tensor reduction\label{symbamp}}

As discussed in Section \ref{prelimi}, the amplitude $i{\cal M}$ factorizes into the 
production of two virtual charged vector bosons
\bea
g(p_1,\lambda_1) + g(p_2,\lambda_2) + W^{-*}(p_3) 
+ W^{+*}(p_4) \to 0\,   \nonumber
\eea
and their decay.  The production amplitude is thus contracted with the vector boson
propagators and lepton currents:
\bea
i{\cal M}(gg\ell\bar\nu\nu'\bar{\ell'}\to 0) = 
\varepsilon_{1\mu_1}\varepsilon_{2\mu_2} 
i{\cal M}^{\mu_1 \mu_2\mu_3\mu_4} P_{\mu_3\nu_3}(p_3) 
P_{\mu_4\nu_4}(p_4) J_{3}^{\nu_3} J_{4}^{\nu_4}\, .  \nonumber
\eea
The coupling constants and colour factors are conveniently absorbed in the 
scattering tensor  ${\cal M}^{\mu_1 \mu_2\mu_3\mu_4}$,
which can be decomposed in terms
of Lorentz tensor structures built from the metric $g^{\mu\nu}$, the external momenta 
$p_1^\mu$, $p_2^\mu$, $p_3^\mu$ and the vector 
$k_0^\mu \equiv\;i\; \epsilon^{\mu p_1 p_2 p_3}$.
Note that $k_0$ products 
are reducible: $k_0^{\mu} k_0^{\nu} = \alpha
g^{\mu\nu} + \beta_{ij} p_i^\mu p_j^\nu$.
The basis of tensor structures that can occur is defined by momentum
conservation, Schou\-ten identities, the transversality/gauge
conditions $\varepsilon_1\cdot p_1=\varepsilon_1\cdot p_2=
\varepsilon_2\cdot p_1=\varepsilon_2\cdot p_2=0$ and current conservation $J_3\cdot p_3=J_4\cdot p_4=0$.
We find
\bea\label{tensordec}
{\cal M}^{\mu_1 \mu_2\mu_3\mu_4} &=& 
        A \, g^{\mu_1\mu_2} g^{\mu_3\mu_4}+\sum\limits_{j_1,j_2,j_3,j_4} C_{j_1 j_2 j_3 j_4}
p_{j_1}^{\mu_1}\,p_{j_2}^{\mu_2} p_{j_3}^{\mu_3}\,p_{j_4}^{\mu_4} \nonumber \\
&&+\sum\limits_{j_3,j_4}  B^1_{j_3 j_4}\,g^{\mu_1\mu_2}\,p_{j_3}^{\mu_3}\,p_{j_4}^{\mu_4}
  +\sum\limits_{j_2,j_4}  B^2_{j_2 j_4}\,g^{\mu_1\mu_3}\,p_{j_1}^{\mu_2}\,p_{j_4}^{\mu_4}
  +\sum\limits_{j_2,j_3}  B^3_{j_2 j_3}\,g^{\mu_1\mu_4}\,p_{j_2}^{\mu_2}\,p_{j_3}^{\mu_3}\nonumber \\
&&+\sum\limits_{j_1,j_4}  B^4_{j_1 j_4}\,g^{\mu_2\mu_3}\,p_{j_1}^{\mu_1}\,p_{j_4}^{\mu_4} 
  +\sum\limits_{j_1,j_3}  B^5_{j_1 j_3}\,g^{\mu_2\mu_4}\,p_{j_1}^{\mu_1}\,p_{j_3}^{\mu_3} 
  +\sum\limits_{j_1,j_3}  B^6_{j_1 j_2}\,g^{\mu_3\mu_4}\,p_{j_1}^{\mu_1}\,p_{j_2}^{\mu_2}
 \nonumber \\ 
 &&+\sum\limits_{j_2} E^1_{j_2 0 0} k_0^{\mu_1} g^{\mu_3 \mu_4} \, p_{j_2}^{\mu_2}
 +\sum\limits_{j_1} E^2_{j_1 00} k_0^{\mu_2} g^{\mu_3 \mu_4} \, p_{j_1}^{\mu_1} \nonumber \\
 &&+\sum\limits_{j_4} E^3_{00 j_4} k_0^{\mu_3} g^{\mu_1 \mu_2} \, p_{j_4}^{\mu_4}
 +\sum\limits_{j_3} E^4_{00 j_3} k_0^{\mu_4} g^{\mu_1 \mu_2} \, p_{j_3}^{\mu_3}\nonumber \\ 
 &&+\sum\limits_{j_2,j_3,j_4} E^1_{j_2 j_3 j_4} k_0^{\mu_1}\,
 p_{j_2}^{\mu_2}\, p_{j_3}^{\mu_3} \, 
 p_{j_4}^{\mu_4}
 +\sum\limits_{j_1,j_3,j_4} E^2_{j_1 j_3 j_4} k_0^{\mu_2}\,
 p_{j_1}^{\mu_1}\, p_{j_3}^{\mu_3} \, 
 p_{j_4}^{\mu_4} \nonumber \\
 &&+\sum\limits_{j_1,j_2,j_4} E^3_{j_1 j_2 j_4} k_0^{\mu_3}\,
 p_{j_1}^{\mu_1}\, p_{j_2}^{\mu_2} \, 
 p_{j_4}^{\mu_4} 
   +\sum\limits_{j_1,j_2,j_3} E^4_{j_1 j_2 j_3} k_0^{\mu_4}\,
   p_{j_1}^{\mu_1}\, p_{j_2}^{\mu_2} \, 
 p_{j_3}^{\mu_3}\,
\eea
with $j_1,j_2= 3$ and $j_3,j_4 \in \{1,2\}$.
Based on Eq.~(\ref{tensordec}) a gauge invariant
representation with 36 coefficients can be derived.
The terms involving an epsilon tensor, i.e.~$k_0$, are parity odd 
and their coefficients are proportional 
to $(m_t^2-m_b^2)$ and vanish if weak isospin is conserved 
\cite{Glover:1988fe,Kao:1990tt}, in particular for massless quarks.

The amplitude is invariant under exchange of the gluons (Bose
symmetry). Since the amplitude is also CP invariant,\footnote{
Note that lepton flavour cannot be 
distinguished in the limit of massless leptons.}
only two helicity amplitudes are independent:
\bea 
{\cal M}^{--J_3J_4}(s,t,u,s_3,s_4) &=& {\cal M}^{++J_4J_3}(s,u,t,s_4,s_3) \,,\label{ppmm}\\ 
{\cal M}^{-+J_3J_4}(s,t,u,s_3,s_4) &=& {\cal M}^{+-J_4J_3}(s,u,t,s_4,s_3) \,,
\eea
with ${\cal M}^{\lambda_1 \lambda_2 J_3J_4} \equiv \varepsilon_{1\mu}^{\lambda_1}
\varepsilon_{2\nu}^{\lambda_2}{\cal M}^{\mu\nu\rho\sigma}J_{3\rho}J_{4\sigma}$. 

Further simplification can be achieved by expressing the helicity amplitudes 
\bea
{\cal M}^{\lambda_1 \lambda_2 J_3J_4} &=& \sum_{j=1}^9 
{\cal C}_j^{\lambda_1\lambda_2}(s,t,u,s_3,s_4,m_b^2,m_t^2) \; \tau_j(J_3,J_4)\,
\nonumber
\eea
in terms of the nine gauge-independent scalar structures
\bea
\tau_j \in \{ J_3\cdot J_4, \: J_3\cdot p_l \; J_4\cdot p_k,\:
J_3\cdot p_l\; J_4\cdot k_0,\: 
J_3\cdot k_0\; J_4\cdot p_k \}\:,
\nonumber
\eea
where $l,k\in\{1,2\}$.
The coefficients ${\cal C}_j^{\lambda_1\lambda_2}$ are linear
combinations of the amplitude coefficients defined in
Eq.~(\ref{tensordec}) and will subsequently be expressed in terms of basis integrals.  
The final coefficients contain negative powers
of the Gram determinant 
\[ \det G = 2 s (tu-s_3s_4)\ , \]
which emerges during
tensor reduction.  In reference frames with back-to-back gluons, it 
is related to the transverse momentum of the $W$ boson: 
$\det G = 2s^2\,p_{3T}^2$.
As the transverse momentum of the vector boson approaches zero, 
the inverse Gram determinant diverges while the amplitude remains finite.
In this phase space region large numerical cancellations occur and
finite machine precision can lead to instabilities during evaluation.
To mitigate this effect, our goal is to cancel as many powers of 
$\det G$ as possible.  For this purpose, we expose the Gram determinant 
in our expressions by introducing the auxiliary vector 
\be
\tilde p_3 = \sqrt{\frac{\det G}{2s^2}}\ (0,\sin\phi_3,\cos\phi_3,0)^T\ .
\nonumber
\ee
For example, if $p_3$ is replaced by $\tilde p_3$ in Eq.~(\ref{gram_det_danger}) the
Gram determinant in the denominator cancels explicity.
More generally, we write the helicity amplitudes 
\bea
\label{amp_pp_tilde}
{\cal M}^{\lambda_1 \lambda_2 J_3J_4} &=& \sum_{j=1}^9 
{\cal \widetilde C}_j^{\lambda_1\lambda_2}(s,t,u,s_3,s_4,m_b^2,m_t^2) 
\; \tilde\tau_j(J_3,J_4) \,
\eea
in terms of the following scalar structures (see Appendix \ref{p3tildeformulas}):
\bea
\tilde\tau_j &\in& 
\{ 
J_3\cdot J_4,\; 
p_1\cdot J_3\,  p_1\cdot J_4, \;  
p_1\cdot J_3\, \tilde p_3\cdot J_4, \;
p_1\cdot J_4\, \tilde p_3\cdot J_3,\;
\tilde p_3\cdot J_3\, \tilde p_3\cdot J_4,\;
\epsilon(p_1,p_2,J_3,J_4),\nonumber \\&&
p_1\cdot J_3\,\epsilon(p_1,p_2,\tilde p_3,J_4),\;
p_1\cdot J_4\,\epsilon(p_1,p_2,\tilde p_3,J_3),\;
\tilde p_3\cdot J_3\,\epsilon(p_1,p_2,\tilde p_3,J_4)\}\, .
\nonumber
\eea

The coefficients in Eq.~(\ref{amp_pp_tilde}) involve tensor (loop momentum)
integrals, which can be written in terms of Lorentz tensors with 
coefficients involving only scalar integrals.  As an alternative to
the standard methods of Refs.~\cite{'tHooft:1978xw,Passarino:1978jh},
we also applied the improved reduction formalism of 
Refs.~\cite{Binoth:2005ff,Binoth:2006rc,Binoth:1999sp} to calculate the 
amplitude.  Here, for instance, a rank two 4-point tensor integral is
reduced via
\bea
I_4^{\mu\nu}(r_1,r_2,r_3,m_1^2,m_2^2,m_3^2,m_4^2) 
= B^{4,2} g^{\mu\nu} + \sum\limits_{j_1,j_2=1}^{3} A^{4,2}_{j_1j_2}
r_{j_1}^{\mu} r_{j_2}^{\nu}\, ,
\nonumber
\eea
where $r_1=p_1$, $r_2=p_1+p_2$ and $r_3=-p_4$ with the external momenta
$p_1,p_2,p_3,p_4$.  The form factors $B^{4,2}$ and $A^{4,2}_{j_1j_2}$
depend on scalar integrals and are ultimately functions of the 
Lorentz invariants $s_{ij}=(p_i+p_j)^2$, $s_{j}=p_j^2$ and the internal masses.
In general, at most rank three tensor box integrals can occur.\footnote{
Analytical and numerical representations for the form factors 
are provided in Ref.~\cite{Binoth:2005ff}.}
For the calculation at hand, we generate explicit analytical representations
in terms of the scalar integral basis\footnote{
The tadpole integral does not appear in this list,
as it can be viewed as a degenerate 2-point integral.}
\bea
I_k \in \{ I_4^{6},I_3^n,I_2^n,1 \}\, .
\nonumber
\eea
In total 27 different scalar integrals appear.
As analytical results do not exist for all 6-dimensional 
4-point functions $I_4^{6}$, we have represented them 
in terms of 3- and 4-point functions in $n=4$.\footnote{This
introduces an inverse Gram determinant, but the expression 
can be grouped such that the combination of scalar
integrals tends to zero as the inverse Gram determinant diverges
and no additional stability problem is introduced.}

Since the process is loop induced, no real corrections exist and 
even for massless internal quarks each loop diagram is infrared (IR) finite.  
We confirmed that the coefficients of IR divergent basis functions vanish.
Since no $ggWW$ counter term exists, the amplitude is also 
ultraviolet (UV) finite.  Note however that each box diagram is UV 
divergent and only the gauge invariant sum of all box graphs
is finite.  We therefore employ dimensional regularization to define the 
amplitudes for individual graphs and evaluate the 2-point function 
coefficients to order $\epsilon= (n-4)/2$ for space-time dimension $n$.
The $V-A$ coupling of the charged vector bosons to the internal quarks
requires a prescription to treat $n$- and 4-dimensional objects consistently. 
We apply standard dimension splitting rules \cite{Veltman:1988au} 
(see Appendix \ref{dimsplit}).

The coefficients in Eq.~(\ref{amp_pp_tilde}) can now be written in terms of basis functions $I_k$ as
\bea
{\cal \widetilde C}_j^{\lambda_1\lambda_2}(s,t,u,s_3,s_4,m_b^2,m_t^2) 
= \sum\limits_{k,l} {\cal \widetilde C}_{jkl}^{\lambda_1\lambda_2}(s,t,u,s_3,s_4,m_b^2,m_t^2)  I_k \, ,
\nonumber
\eea
where the coefficient ${\cal \widetilde C}_{jkl}^{\lambda_1\lambda_2}$ corresponds to diagram 
$l$ and is a rational polynomial that is computed with and saved as {\sc Form} 
\cite{Vermaseren:2000nd} code.\footnote{
Unevaluated amplitude expressions for each contributing Feynman graph 
have been compared with output from FeynArts 3.2 \cite{Hahn:2000kx}.
}
The irreducible amplitude coefficients are then simplified with {\sc Maple}.  
First, each ${\cal \widetilde C}_{jkl}^{\lambda_1\lambda_2}\,\tilde\tau_j$ expression
is simplified.  For $\lambda_1\lambda_2=++,--$ all inverse Gram determinants cancel 
in this step.  For $\lambda_1\lambda_2=+-,-+$, however, one inverse power survives.  
Next, we sum over the diagrams, which facilitates further simplification,
because discrete symmetries and gauge invariance are restored.
The final output is converted to {\sc Fortran} code. All steps are automatized.

For the triangle topologies with Higgs exchange we thus obtain the well-known result
\bea
i{\cal M}(gg\to H \to W^{+*}W^{-*}\to\ell\bar\nu\nu'\bar{\ell'}) = {\cal M}^{\lambda_1\lambda_2 J_3J_4} \frac{1}{s-m_H^2+i\, m_H \Gamma_H}\,
\nonumber
\eea
with 
\bea
{\cal M}^{\lambda_1\lambda_2 J_3J_4} &=& {\cal \widetilde C}_H^{\lambda_1
  \lambda_2}(s,m_b^2,m_t^2) \: 
J_3\cdot J_4 \, 
\nonumber
\eea
and
\bea
{\cal \widetilde C}_H^{\pm\pm}(s,m_b^2,m_t^2) &=&   
 - 2 \,  m_t^2 \,(s-4\, m_t^2)\, I_3^n(s,0,0,m_t^2,m_t^2,m_t^2) 
\nonumber\\&& 
     - 2 \,  m_b^2 \,(s-4\, m_b^2)\, I_3^n(s,0,0,m_b^2,m_b^2,m_b^2)
     + 4 \, ( m_t^2 + m_b^2 )
\,,\nonumber\\
{\cal \widetilde C}_H^{\pm\mp}(s,m_b^2,m_t^2) &=&   0\,.
\nonumber
\eea
Here, only the helicity combinations $\lambda_1\lambda_2$ $=++$ and $--$ contribute,
since the intermediate scalar forces the gluons to be in an $L=S=0$ state.
Eq.~(\ref{ppmm}) implies ${\cal \widetilde C}_H^{++}={\cal \widetilde C}_H^{--}$.
The explicit results for the box topologies are too complex to be presented.

Two independent calculations of the amplitude have been performed.
All symmetry relations of the amplitude have been checked analytically.
In one calculation the tensor reduction methods described in this section
were applied, while standard methods \cite{'tHooft:1978xw,Passarino:1978jh} were used 
in the other.  

\subsection{Numerical amplitude evaluation \label{numamp}}

The symbolic evaluation method described in Section \ref{symbamp}
strongly reduces the destabilizing effects of inverse Gram determinants in the
final amplitude representation, but does not completely remove them for the $+-$ and 
$-+$ helicity combinations.  
When evaluated in double precision, our analytic expression 
for the amplitude thus exhibits numerical instabilities in the extreme forward 
scattering region, e.g. $p_T(W^\pm) \to 0$, where $(\det G)^{-1}$ diverges.  
Since the $\nu$ pair is not detected, this phase space region still 
contributes to the cross section after application of the selection cuts.

Numerical instabilities can be remedied by evaluating the amplitude 
in quadruple precision. But, a huge runtime penalty is incurred 
in comparison to double precision. In order to overcome this practical
problem, one can restrict the use of quadruple precision to a small
region in phase space where
\begin{displaymath}
p_T(W^\pm) < 6\ \text{GeV} 
\end{displaymath}
and
\begin{displaymath}
p_T(W^\pm) < 1\ \text{GeV}\quad \text{or}\quad 
\max\left(\left|\sqrt{p^2_{W^\pm}} - M_W\right|\right) > 5\,\Gamma_W\,, 
\end{displaymath}
while double precision is used in the remainder of the phase space.
Using this ``mixed'' mode of our numerical program, the results
presented below in Section~\ref{results-section} were calculated with
no indication of numerical instabilities and no significant runtime
overhead.
For a specific phase space configuration we compared numerical results for 
$|{\cal M}|^2$ obtained with our independent amplitude calculations 
and found agreement.
We use LoopTools \cite{Hahn:1998yk} to evaluate the scalar integrals numerically. 

\subsection{Cross-section calculation and the \texttt{GG2WW} program \label{sec:gg2ww}}

The cross sections and distributions presented in Section~\ref{results-section} 
were verified with two independent phase space and Monte Carlo integration 
implementations.

Our public program, named \texttt{GG2WW}, includes all background and signal
contributions, full spin correlations, off-shell and interference
effects, as well as finite top and bottom quark mass effects.  It can
be used either at the parton level or to generate weighted or
unweighted events in Les Houches Accord format
\cite{LesHouchesEventGeneratorInterface}.  A combination of the
multi-channel \cite{Berends:1994pv,Kleiss:1994qy} and phase
space-decomposition \cite{Kauer:2001sp,Kauer:2002sn} Monte Carlo
integration techniques was used with appropriate mappings to
compensate peaks in the amplitude.  In addition, automatized
VEGAS-style \cite{VEGAS} adaptive sampling is employed using
\texttt{OmniComp-Dvegas}, which features a parallel mode (including
histogram filling) \cite{OmniComp}.  Parton distribution functions are
included via the LHAPDF package \cite{LHAPDF}.  Selection cuts and
histograms can be specified in a user-friendly format.  The program is
available on the Web \cite{GG2WW} and has already been used by ATLAS
and CMS in recent $H\to WW$ studies
\cite{Buttar:2006zd,Drollinger:2005ug,CMS-TDR}.

\section{Results \label{results-section}}

In this section we present numerical results for the process $pp \to
W^{\ast}W^{\ast}\to \ell\bar{\nu}\bar{\ell'}\nu'$ at the LHC.  We
tabulate the total cross section and the cross section for two sets of
experimental cuts.  We focus on the impact of the massive top-bottom
loop, which has been neglected in~Ref.~\cite{Binoth:2005ua}, and the size
of the signal-background interference. As discussed in
Section~\ref{sec:gg2ww}, we provide a public parton-level event
generator for the process $gg \to W^{\ast}W^{\ast} \to {\rm
  leptons}$~\cite{GG2WW}, which can be used to study alternative sets
of cuts or to generate any kind of distribution.
 
The experimental cuts include a set of ``standard
cuts''~\cite{Dixon:1999di}, where we require both charged leptons to
be produced at $p_{T,\ell} > 20$ GeV and $|\eta_\ell| < 2.5$
motivated by detector coverage, and a
missing transverse momentum $\sla{p}_T > 25$~GeV characteristic for 
leptonic $W$ decays. Cross sections
calculated with this set of cuts will be labeled $\sigma_{std}$.
Various further cuts have been proposed for the experimental Higgs
searches to enhance the signal-to-background
ratio~\cite{Dittmar:1996ss,Dittmar:1996sp,DittmarDreinerIII,
  JakobsTrefzger,Green:2000um, Davatz:2004zg}.  As in our previous
publication~\cite{Binoth:2005ua} we have studied a set of cuts similar
to those advocated in a recent experimental
study~\cite{Davatz:2004zg}.  In addition to the ``standard cuts''
defined above, we require that the opening angle between the two
charged leptons in the plane transverse to the beam direction should
satisfy $\Delta\phi_{T,\ell\ell} < 45^{\circ}$ and that the dilepton
invariant mass $M_{\ell\ell}$ be less than $35$~GeV.  Furthermore, the
larger and smaller of the charged lepton transverse momenta are restricted as
follows: $25~\mbox{GeV} < p_{T,{\rm min}}$ and $35~\mbox{GeV} < p_{T,
  {\rm max}} < 50~\mbox{GeV}$.  Finally, a jet veto is imposed that
removes events with jets where $p_{T,{\rm jet}} > 20$~GeV and
$|\eta_{\rm jet}| < 3$. Cross sections evaluated with the Higgs
selection cuts will be labeled $\sigma_{bkg}$.

To obtain numerical results we have used the same set of input parameters
as in~Ref.~\cite{Binoth:2005ua}:
\begin{displaymath}
\begin{array}{llllll}
M_W & = 80.419~{\rm GeV}, \quad &
M_Z & = 91.188~{\rm GeV}, \quad &
G_\mu & = 1.16639 \times 10^{-5}~{\rm GeV}^{-2}, \\
\Gamma_W & = 2.06~{\rm GeV}, &
\Gamma_Z & = 2.49~{\rm GeV}, &
V_{\rm CKM} & = \mathbbm{1}\, .
\end{array}
\end{displaymath}
The top-bottom quark-loop contribution has been evaluated using
$M_t = 178~{\rm GeV}$ and $M_b = 4.4~{\rm GeV}$. To study the
signal-background interference we have chosen three Higgs mass values
($M_H$ = 140, 170, 200 GeV), with the corresponding Higgs widths
$\Gamma_H = 0.008235~{\rm GeV}$ $(M_H = 140~{\rm GeV})$, $\Gamma_H =
0.3837~{\rm GeV}$ $(M_H = 170~{\rm GeV})$, and $\Gamma_H = 1.426~{\rm
  GeV}$ $(M_H = 200~{\rm GeV})$, as calculated by
HDECAY~\cite{Djouadi:1997yw}.  The weak mixing angle is given by
$c_{\rm w} = M_W/M_Z,\ s_{\rm w}^2 = 1 - c_{\rm w}^2$, and the
electromagnetic coupling has been defined in the $G_\mu$ scheme as
$\alpha_{G_\mu} = \sqrt{2}G_\mu M_W^2s_{\rm w}^2/\pi$. The masses of
external fermions have been neglected.  The $pp$ cross sections have
been calculated at $\sqrt{s} = 14$~TeV employing the
LHAPDF~\cite{LHAPDF} implementation of the CTEQ6L1 and CTEQ6M
\cite{Pumplin:2002vw} parton distribution functions at tree- and
loop-level, corresponding to $\Lambda^{\rm LO}_5 = 165$ MeV and
$\Lambda^{\overline{{\rm MS}}}_5 = 226$ MeV with 1- and 2-loop
running for $\alpha_s(\mu)$, respectively.\footnote{We observe a
  relative deviation of ${\cal O}(10^{-4})$ when comparing cross
  sections obtained with the LHAPDF implementation of CTEQ6 to those
  obtained with the original CTEQ6 implementation.} The
renormalization and factorization scales are set to $M_W$.
Fixed-width Breit-Wigner propagators are used for unstable gauge
bosons.

In Table~\ref{tbl:xsections} we present the total cross section and
the cross section for two sets of experimental cuts: ``standard cuts''
($\sigma_{std}$ ) and Higgs search cuts ($\sigma_{bkg}$) as defined
above. The results for the gluon-fusion cross section in
Table~\ref{tbl:xsections} include the contribution from the massive
top-bottom loop and supersede our previous
calculation~\cite{Binoth:2005ua}, which was based on intermediate
light quarks only. For reference, we also show the LO and NLO quark
scattering cross sections, which have been computed with
MCFM~\cite{Campbell:1999ah}.  As already demonstrated
in~Ref.~\cite{Binoth:2005ua} the $gg$ process only yields a 5\% correction
to the total $WW$ cross section calculated from quark scattering at
NLO QCD.  When realistic Higgs search selection cuts are applied the
correction increases to 30\%.  For a discussion of the renormalization
and factorization scale uncertainties we refer
to~Ref.~\cite{Binoth:2005ua}.

\TABLE{
\renewcommand{\arraystretch}{1.5}
\hspace*{0cm}
\begin{tabular}{|c|c|c c|c|c|}
 \cline{2-6}
\multicolumn{1}{c|}{} & \multicolumn{5}{c|}{$\sigma(pp \to W^{\ast}W^{\ast}\to
   \ell\bar{\nu}\bar{\ell'}\nu')$~[fb],\ \ LHC} \\ \cline{2-6}
\multicolumn{1}{c|}{}  & & \multicolumn{2}{c|}{\raisebox{1ex}[-1ex]{$q\bar{q}$} } & & \\[-1.5ex]
\cline{3-4} 
\multicolumn{1}{c|}{}  & \raisebox{3ex}[-3ex]{$gg$} &
\raisebox{0.7ex}{LO} & \raisebox{0.7ex}{NLO} &
\raisebox{2.75ex}[-2.75ex]{$\frac{\sigma_{\rm NLO}}{\sigma_{\rm LO}}$} & 
  \raisebox{2.75ex}[-2.75ex]{$\frac{\sigma_{{\rm NLO}+gg}}{\sigma_{\rm NLO}}$}
\\[-1.5ex]
\hline
 $\sigma_{tot}$ & 60.00(1) & $875.8(1)$ &
 $1373(1)$ & $1.57$ & 1.04 \\
 \hline
 $\sigma_{std}$ & 29.798(6) & $270.5(1)$ &
 $491.8(1)$ &  $1.82$ & 1.06 \\
 \hline
 $\sigma_{bkg}$ & 1.4153(3) & $4.583(2)$ &
 $4.79(3)$ &  $1.05$ & 1.30 \\
 \hline
\end{tabular}\\[.4cm]
\caption{\label{tbl:xsections}
  Cross sections for the gluon and quark scattering contributions to
  $pp \to W^{\ast}W^{\ast}\to \ell\bar{\nu}\bar{\ell'}\nu'$ at the LHC
  ($\sqrt{s} = 14$ TeV) without selection cuts ($tot$), with standard
  LHC cuts ($std$: $p_{T,\ell} > 20$ GeV, $|\eta_\ell| < 2.5$,
  $\sla{p}_T > 25$ GeV) and Higgs search selection cuts ($bkg$)
  applied. The Higgs signal has not been included. The integration
  error is given in brackets. We also show the ratio of the NLO to LO
  quark scattering cross sections and the ratio of the combined
  NLO+$gg$ contribution to the NLO cross section.}}

The importance of the top-bottom loop contribution can be inferred
from Table~\ref{tbl:23interference}, where we compare the results
based on intermediate light quarks of the first two generations
only~\cite{Binoth:2005ua} to the contribution of the top-bottom loop
and the interference between massless and massive quark loops. We find
that the top-bottom loop increases the theoretical prediction by 12\%
and 15\% for the inclusive cross section, $\sigma_{tot}$, and the
cross section with standard cuts, $\sigma_{std}$, respectively.  After
imposing Higgs search cuts, however, the contribution of the massive
quark loop is reduced to 2\% only, which is almost entirely due to
interference with the massless loop amplitude.  The reduction can largely be
attributed to the cut on $\Delta\phi_{T,\ell\ell}$ as can be seen in
Fig.~\ref{fig:delphill}: while the impact of the top-bottom loop is
sizeable in most of phase space, it is strongly reduced in the region
$\Delta\phi_{T,\ell\ell} < 45^{\circ}$ selected by the Higgs search
cuts.
In Fig.~\ref{fig:delphill}, we also see that off-shell effects slightly 
decrease the cross section and become negligible for almost back-to-back 
charged leptons.  Allowing for arbitrary invariant masses of the $W$ bosons 
changes the complete background cross section with standard cuts
by --2.8\%, which increases to --6.4\% when Higgs search cuts are applied.

\EPSFIGURE[p] {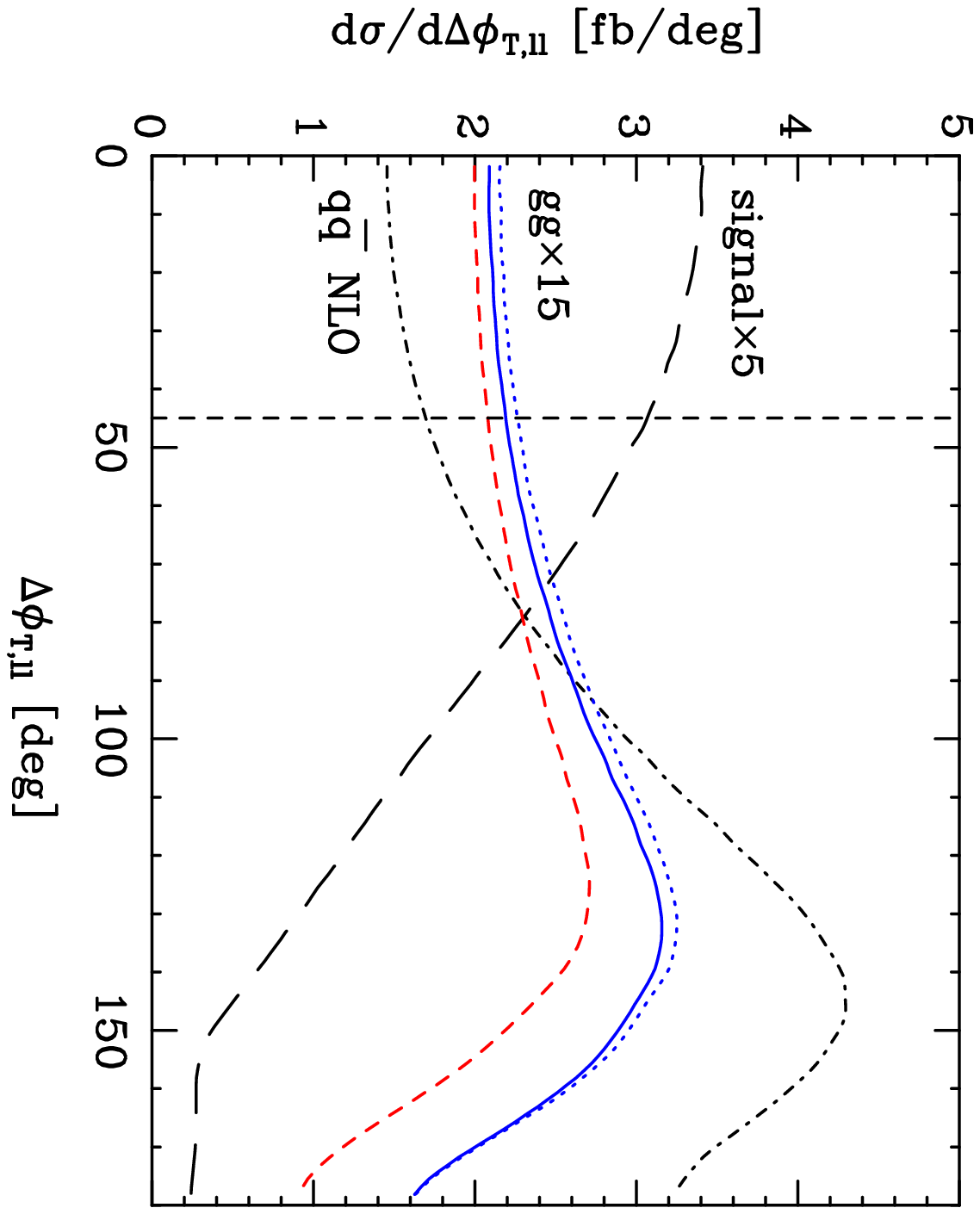,width=8cm,angle=90}
{\label{fig:delphill} Distribution in the transverse-plane opening
  angle of the charged leptons $\Delta\phi_{T,\ell\ell}$ for $pp \to
  W^{\ast}W^{\ast}\to \ell\bar{\nu}\bar{\ell'}\nu'$ at the LHC.
  Displayed is the background process from gluon scattering without (dashed) 
  and with the top-bottom loop (solid) -- the latter also without off-shell $W$ 
  effects (dotted) -- each multiplied
  with a factor 15; the background from quark scattering at NLO
  (dot-dashed); and the signal process for $M_H = 170$~GeV (long-dashed)
  multiplied with a factor 5. Input parameters as defined in the main
  text.  Standard LHC cuts have been applied. The Higgs search cuts
  select the region $\Delta\phi_{T,\ell\ell} < 45^{\circ}$ left of the
  vertical dashed line.}

\TABLE{
\renewcommand{\arraystretch}{1.5}
\hspace*{0cm}
\begin{tabular}{|c|c|c|c|c|}
 \cline{2-5}
\multicolumn{1}{c|}{} & \multicolumn{4}{c|}{$\sigma(pp \to W^{\ast}W^{\ast}\to
   \ell\bar{\nu}\bar{\ell'}\nu')$~[fb],\ \ LHC} \\ \cline{2-5}
\multicolumn{1}{c|}{}  & quark loop & quark loop & & interference\\[-1.5ex]
\multicolumn{1}{c|}{}  &  
generations 1,2  & 
generation 3 &
\raisebox{1.75ex}[1.75ex]{$\frac{\displaystyle \rm gen.\, 1,2,3}{\displaystyle \rm gens.\, 1,2}$}
&
$\frac{\rm gens.\, 1,2,3}{\rm [gens.\, 1,2] + [gen.\, 3]}$
 \\[0.5ex]
\hline
 $\sigma_{tot}[gg]$ & 53.64(1) & 2.859(3) & 1.12 & 1.06 \\
 \hline
 $\sigma_{bkg}[gg]$ & 1.3837(3) & 0.00377(2) & 1.02 & 1.02 \\
 \hline
\end{tabular}\\[.4cm]
\caption{\label{tbl:23interference} 
  Cross sections for the gluon scattering contribution to $pp \to
  W^{\ast}W^{\ast}\to \ell\bar{\nu}\bar{\ell'}\nu'$ at the LHC
  ($\sqrt{s} = 14$ TeV) without selection cuts ($tot$) and Higgs
  search selection cuts ($bkg$) applied.  We show the cross section
  with 2 massless generations~\protect\cite{Binoth:2005ua}, the
  contribution of the top-bottom loop, and the size of the
  interference effects. The Higgs signal has not been included. The
  integration error is given in brackets.}}

An interesting distribution, which we did not display
in~Ref.~\cite{Binoth:2005ua} is the transverse mass distribution
$\mbox{d}\sigma/\mbox{d}M_T$, where one uses a measurable proxy for
the Higgs transverse mass defined by
$M_T =\sqrt{(E_{T,\ell\ell}+\sla{E}_T)^2 -
  ({\vec{p}}_{T,\ell\ell}+\sla{\vec{p}}_T)^2}$, with $E_{T,\ell\ell} =
\sqrt{p_{T,\ell\ell}^2 + m_{\ell\ell}^2}$ and 
$\sla{E}_T = \sqrt{\sla{p}_T^2 + m_{\ell\ell}^2}$~\cite{Rainwater:1999sd}. 
Cuts on the transverse mass $M_T$ provide an additional handle to suppress the
background with respect to the Higgs signal, see
e.g.~Refs.~\cite{Duhrssen:2005bz,JakobsTrefzger,Green:2000um}. In
Fig.~\ref{fig:mtWW_RZ} we compare the $M_T$-distribution of the
$q\bar{q}$ background in LO and NLO with the contribution from
gluon-gluon scattering before and after applying Higgs search cuts.
The figures reveal that the gluon-induced contribution becomes the
dominant higher-order correction to the background process after Higgs
selection cuts have been imposed.

\FIGURE{
\begin{minipage}[c]{.49\linewidth}
\flushright \includegraphics[height=7.cm, clip=true, angle=90]{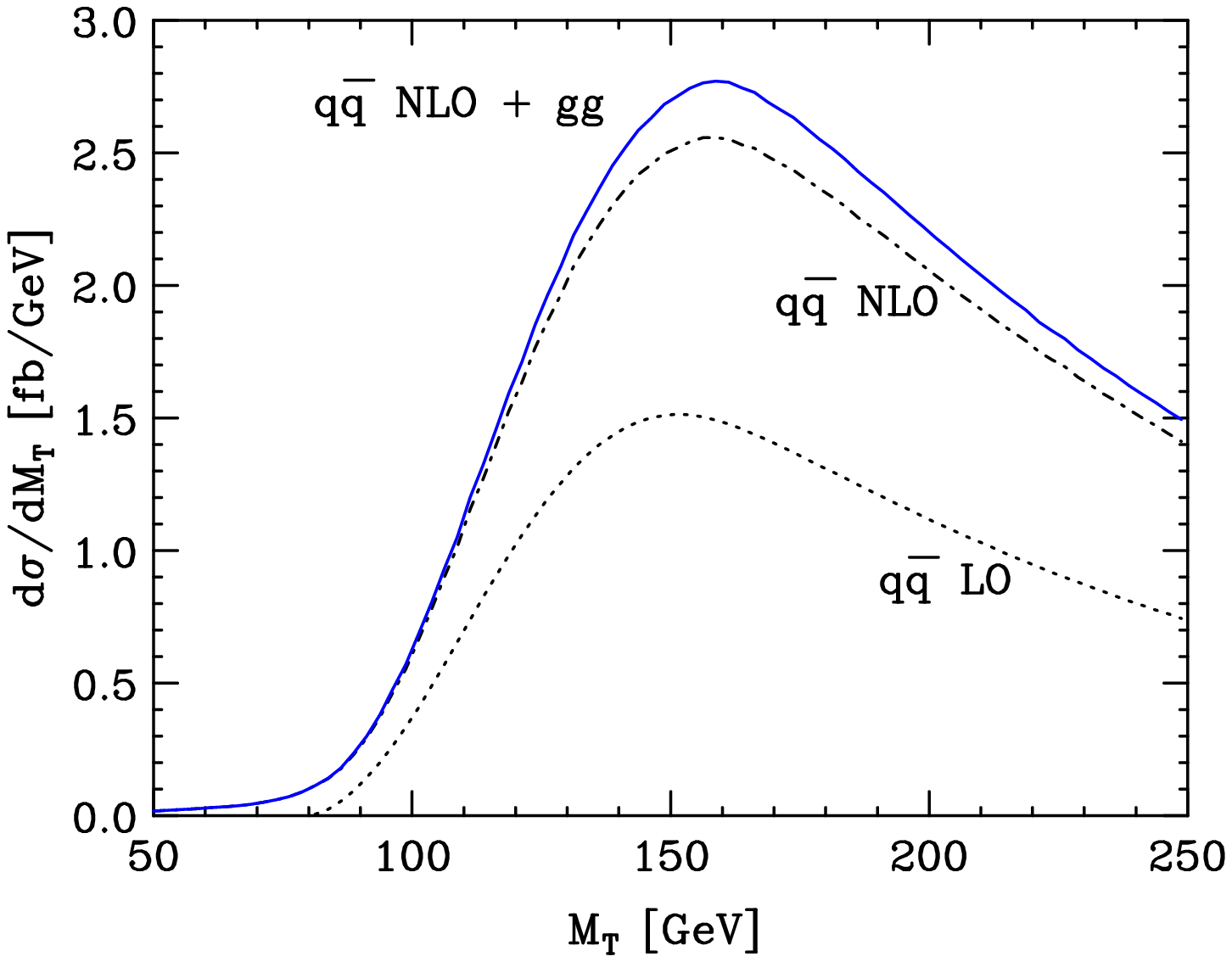}
\end{minipage} \hfill
\begin{minipage}[c]{.49\linewidth}
\flushleft \includegraphics[height=7.cm, angle=90]{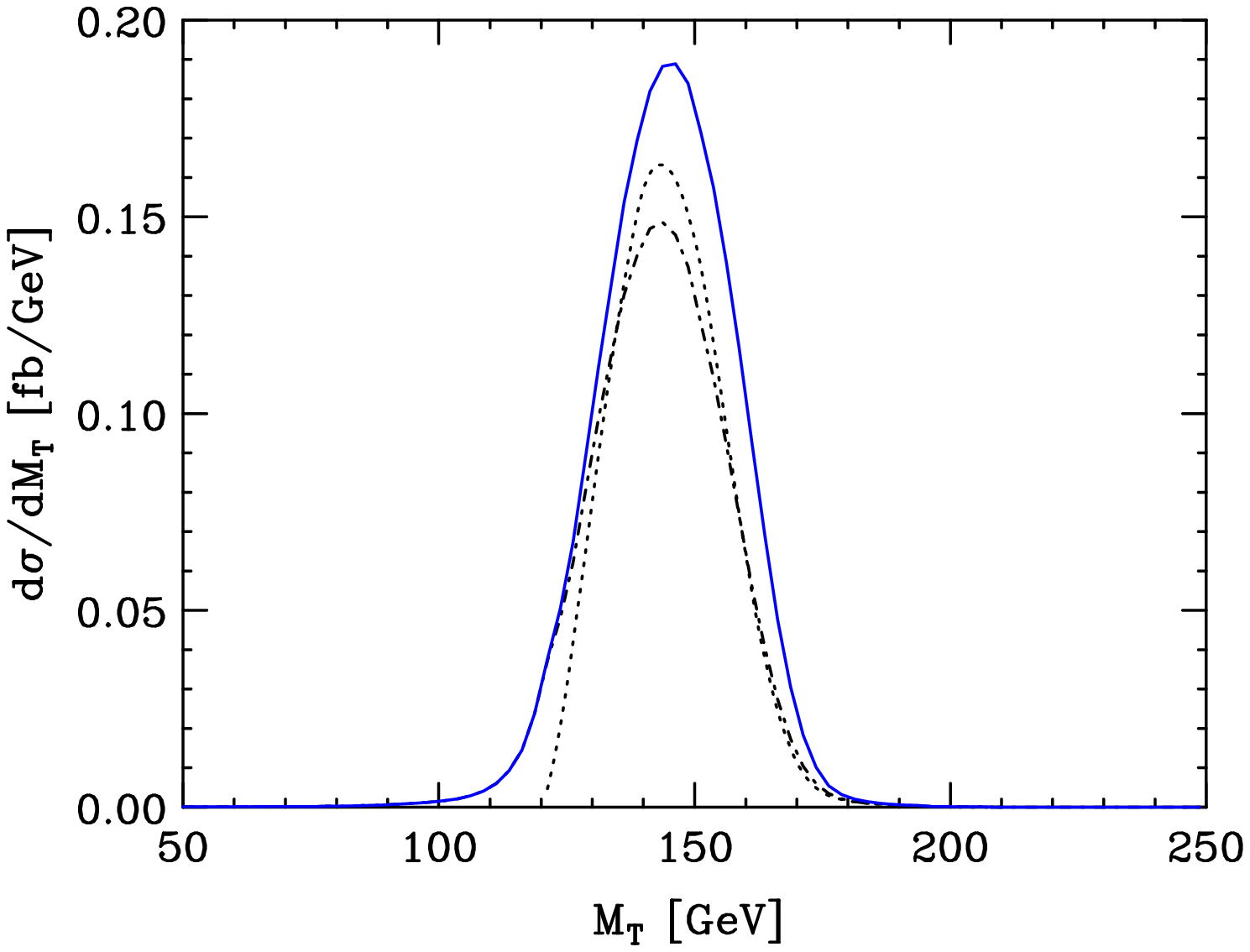}
\end{minipage}\\[0.2cm]
\caption{{\label{fig:mtWW_RZ} Distribution in the transverse mass $M_T$ (as defined in the
    text) with standard cuts (left figure) and Higgs search cuts (right
    figure).  Displayed are the total background from quark scattering
    at NLO and gluon-fusion (solid), and from quark scattering alone
    at LO (dotted) and NLO (dot-dashed).}}}

We now turn to the discussion of the interference effects between the
gluon-gluon induced signal and background processes.
Table~\ref{tbl:interference} shows cross sections for the signal and
gluon-fusion background with and without interference.  We show
results for $M_H = 140$, 170 and 200~GeV, spanning the Higgs mass
range for which the $H\to WW$ decay mode is of particular relevance.
It turns out that the interference effects are quite small and never
exceed 10\% of the gluon-induced signal plus background cross section.
Adding the NLO background contribution from quark scattering, the overall
effect of the interference term is always less than 5\%.  These results
are consistent with the small signal-background interference observed
in $pp\to H\to\gamma\gamma$~\cite{Dicus:1987fk,Dixon:2003yb}.
Due to the imposed jet veto, we expect only small effects at the LHC when 
NLO corrections are taken into account for the signal \cite{Catani:2001cr}
and $gg$ background.

\TABLE{
\renewcommand{\arraystretch}{1.5}
\begin{tabular}{|c|c|c|c||c|c|c|}
 \cline{2-7}
\multicolumn{1}{c|}{} & \multicolumn{6}{c|}{$\sigma[gg (\to H) \to W^{\ast}W^{\ast}\to
   \ell\bar{\nu}\bar{\ell'}\nu']\ \ {\rm [fb]}$} \\
\hline
\multicolumn{1}{|c|}{\rm cut selection} &
\multicolumn{3}{c||}{$tot$} &
\multicolumn{3}{c|}{$bkg$}
\\
\hline
\multicolumn{1}{|c|}{$M_H {\rm [GeV]}$} &
$140$ &
$170$ &
$200$ &
$140$ &
$170$ &
$200$ \\
\hline
$\sigma[{\rm signal}]$  & 79.83(2)  & 116.23(3)  & 75.40(2) & 1.8852(5) & 12.974(2) & 1.6663(7) \\
 \hline
$\sigma[{\rm signal+bkg(gg)}]$ & 132.50(5) & 174.58(9) & 134.46(5) & 3.174(2) & 15.287(6) & 3.413(2)  \\
 \hline
$\frac{\sigma[{\rm sig+bkg(gg)}]}{\sigma[{\rm signal}]+\sigma[{\rm bkg(gg)}]}$
& $0.948$ & $0.991$ & $0.993$ & $0.962$ & $1.062$ & $1.108$ \\[1ex]
 \hline
\end{tabular}\\[0.2cm]
\caption{\label{tbl:interference} 
  Interference effects between the signal and gluon-induced background
  processes.  Details as in Table~\protect\ref{tbl:xsections}.}}

In Fig.~\ref{fig:winvmass140} we finally show the $W^{-}$
invariant-mass distribution without applying selection cuts 
for $M_H = 140~{\rm GeV}\! < 2 M_W$. Although this distribution is not 
measurable it is instructive in understanding interference and
off-shell effects. 
We compare the signal and the
gluon-induced background with and without interference effects.  In
addition to a pronounced, narrow resonance at $M_W$, the $W^{-}$
invariant mass distribution exhibits a second small and broad
resonance at about 55~GeV, which stems from the kinematic constraint
imposed by the dominant Higgs resonance in the
signal process. We note that without any selection cuts and for Higgs
masses around 140~GeV, the gluon-induced background exceeds the
Higgs-boson signal in the resonant region near $M_W$.

\FIGURE{
\begin{minipage}[c]{.49\linewidth}
\flushright \includegraphics[height=7.cm, clip=true, angle=90]{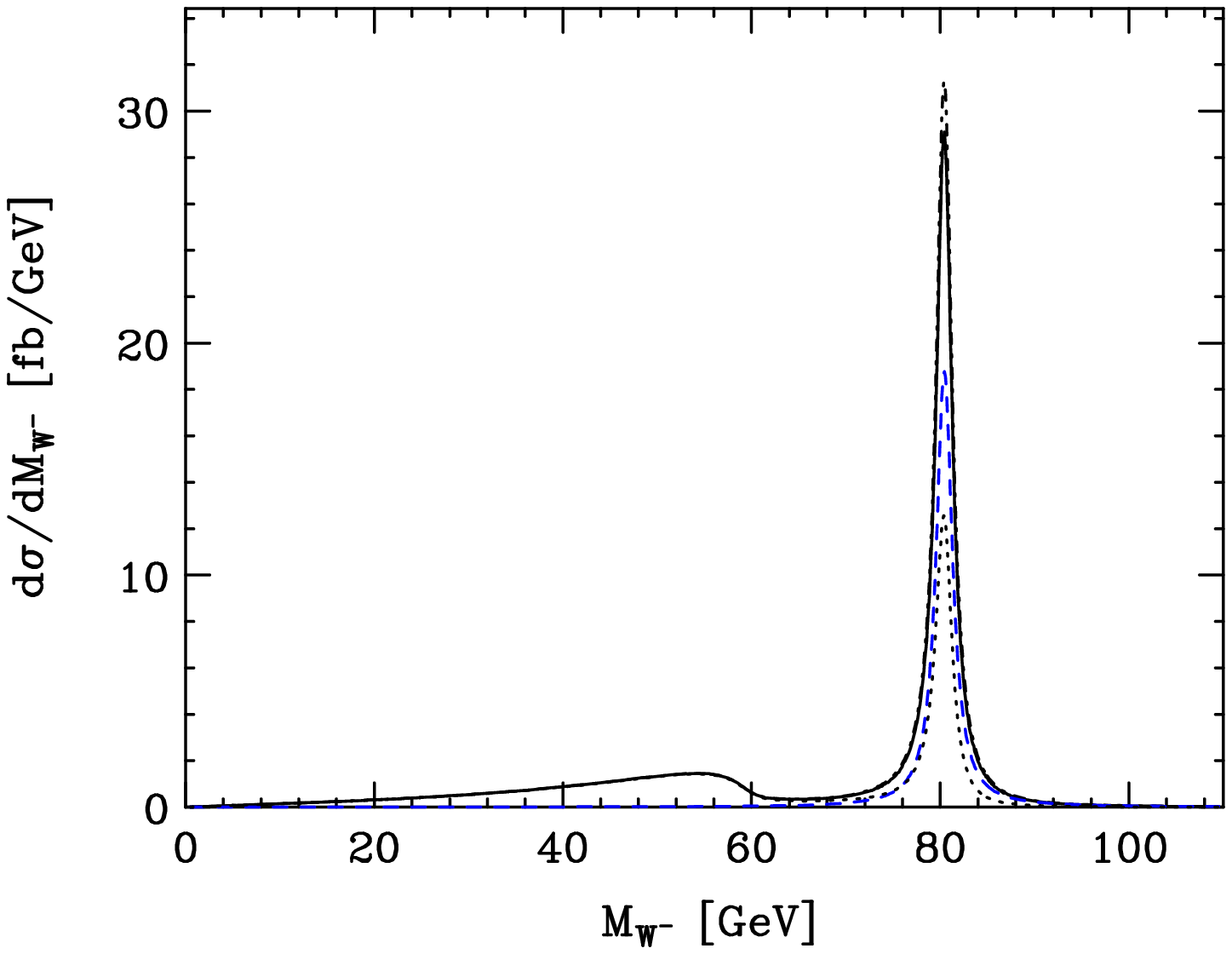}
\end{minipage} \hfill
\begin{minipage}[c]{.49\linewidth}
\flushleft \includegraphics[height=7.cm, angle=90]{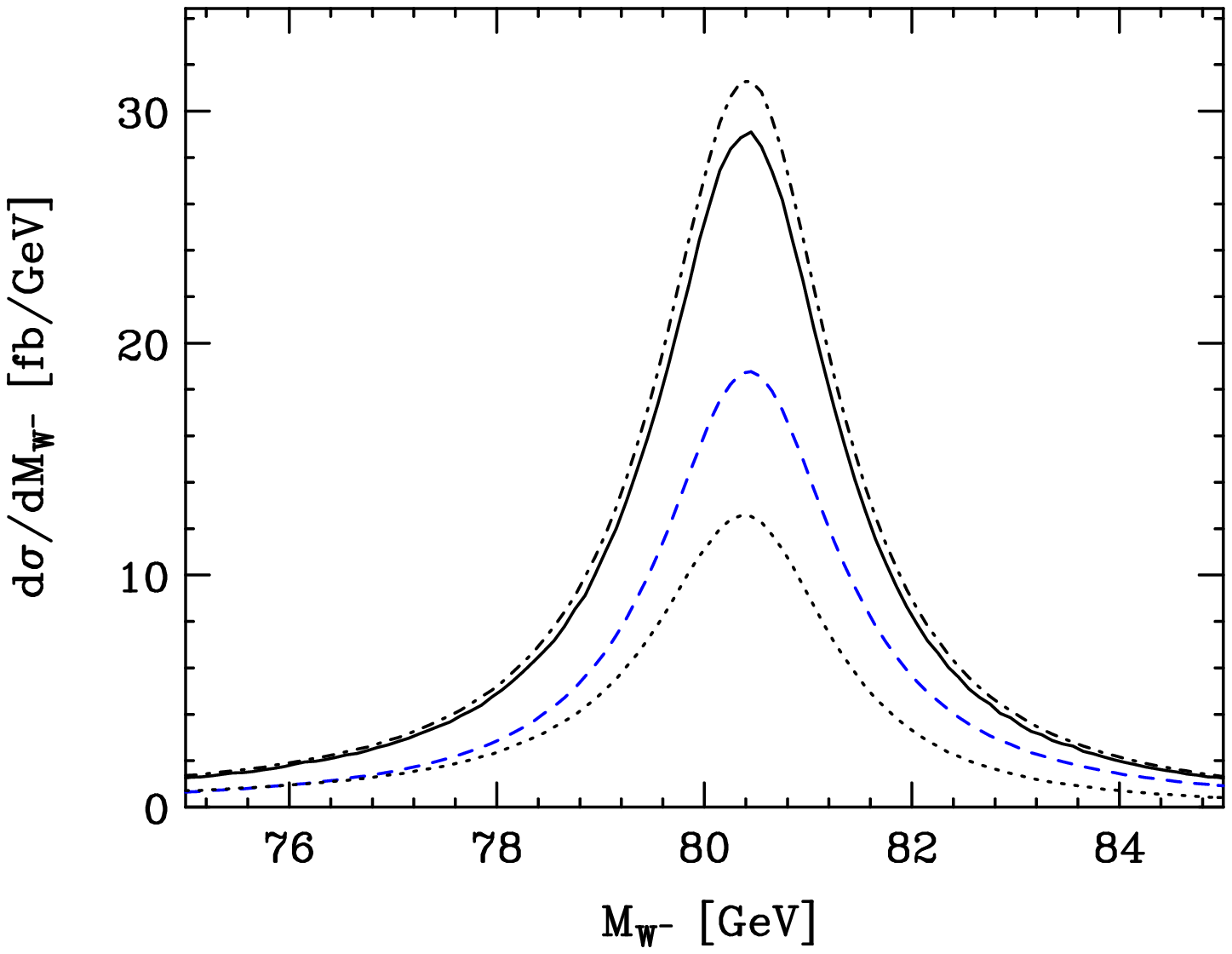}
\end{minipage}\\[0.2cm]
\caption{\label{fig:winvmass140}
  Distribution in the $W^-$ invariant mass for $M_H$ = 140 GeV.  The
  resonant region is magnified on the right hand side.  Complete $gg$
  background without signal (dashed, blue), signal only (dotted),
  signal and background with interference (solid) and signal and
  background without interference (dot-dashed) are displayed.  
  No selection cuts are applied  ($tot$).  } }

\section{Conclusions \label{concl-section}}

We have presented the first complete calculation of the loop-induced
gluon-fusion process $gg \to W^{\ast}W^{\ast} \to
\ell\bar{\nu}\bar{\ell'}\nu'$, including intermediate light and heavy
quarks, and studied its importance for Higgs boson searches in the
$H\to WW$ channel.  We find that the top-bottom loop, which had been
neglected in~Ref.~\cite{Binoth:2005ua} contributes at a level of about
10-15\% to the inclusive gluon-induced cross section but is strongly
suppressed after Higgs search cuts have been imposed.  We have also
studied interference effects between signal and background processes
and found them to be small (about 5\% or less) in the relevant Higgs
mass range between $M_H=140-200$~GeV.  We provide the {\tt GG2WW} package
\cite{GG2WW}, a public parton-level Monte Carlo program and event 
generator
for the process $gg \to W^{\ast}W^{\ast} \to
\ell\bar{\nu}\bar{\ell'}\nu'$ that can be used to calculate cross
sections with any set of cuts or any kind of differential
distribution, or to generate weighted or unweighted events for
experimental analyses.
 
\acknowledgments The work of T.B. and N.K. was supported by the Deutsche
Forschungsgemeinschaft (DFG) under contract number BI 1050/1 and the
Bundesministerium f\"ur Bildung und Forschung (BMBF, Bonn, Germany)
under contract number 05HT1WWA2. T.B. is supported by  
the Particle Physics and Astronomy Research Council (PPARC) of the UK
and the Scottish Universities Physics Alliance (SUPA).

\clearpage

\begin{appendix}
\renewcommand{\theequation}{\Alph{section}.\arabic{equation}}
\setcounter{equation}{0}

\section{Auxilliary vector relations\label{p3tildeformulas}}

\bea
p_3 = \frac{t-s_3}{s} p_1 + \frac{u-s_3}{s} p_2 + \tilde p_3\nonumber\\
p_4 = \frac{u-s_4}{s} p_1 + \frac{t-s_4}{s} p_2 - \tilde p_3 \nonumber\, 
\eea
\begin{gather*}
p_1\cdot \tilde p_3=p_2\cdot \tilde p_3=0 \\
k_0^\mu=\epsilon(p_1,p_2,\tilde p_3,\mu) \\
\tilde p_3^2 = -\frac{\det G}{2\;s^2}
\end{gather*}

To define the basis used in Eq.~(\ref{amp_pp_tilde}) we exploit that
\bea
p_3\cdot J_3 &=& \frac{t-s_3}{s} p_1\cdot J_3 + \frac{u-s_3}{s}
p_2\cdot J_3 + \tilde p_3\cdot J_3 
= 0\nonumber\ ,\\
p_4\cdot J_4 &=& \frac{u-s_4}{s} p_1\cdot J_4 + \frac{t-s_4}{s}
p_2\cdot J_4 - \tilde p_3\cdot J_4 
= 0\nonumber\ ,\\
J_3\cdot J_4 &=& \frac{2}{s} ( p_1\cdot J_3\;  p_2\cdot J_4 + p_2\cdot J_3\;  p_1\cdot J_4 )
                 -\frac{s}{u t - s_3 s_4} \tilde p_3\cdot J_3  \tilde p_3\cdot J_4\nonumber\\&&
                 +\frac{4}{s(u t - s_3 s_4)} \epsilon(p_1,p_2,\tilde
                 p_3, J_3)\, 
\epsilon(p_1,p_2,\tilde p_3, J_4) \nonumber\ ,\\
\epsilon(p_1,p_2,J_3,J_4) &=& -\frac{s}{u t -s_3 s_4}[
                                                       \tilde p_3 \cdot J_3  k_0 \cdot J_4 
                                                    -  \tilde p_3 \cdot J_4  k_0 \cdot J_3 
                                                     ] \ .
\nonumber
\eea

\section{Dimension splitting formulae\label{dimsplit}}

When using dimensional regularization in combination with genuinely 4-dimensional 
objects, one is forced to apply a calculational scheme to deal with the $\gamma_5$ 
problem.  We apply standard dimension splitting rules \cite{Veltman:1988au}
and use an $n$-dimensional loop momentum $k$ and
gamma matrices $\gamma_\mu$, but work with 4-dimensional external momenta.  
The following rules are sufficient to evaluate
the diagrams in Fig.~\ref{graphs}:
\bea
&& k =\hat k + \tilde k \quad , \qquad  k^2 = \hat k^2 + \tilde k^2 \quad ,\nonumber\\
&& \gamma= \hat \gamma + \tilde \gamma \quad , \qquad \{ \hat \gamma^\mu,\tilde \gamma^\nu \} = 0 
\quad , \qquad\{ \hat \gamma^\mu,\hat \gamma^\nu \} = \hat g^{\mu\nu} \quad , \qquad \nonumber\\
&& p_j^\mu \gamma_\mu = p_j^\mu \hat \gamma_\mu \quad , \qquad p_j\cdot k = p_j\cdot \hat k\quad ,\nonumber\\ 
&& \{ \hat\gamma^\mu ,\gamma_5 \} = 0 \quad , \qquad  [ \tilde \gamma^\mu ,\gamma_5 ] = 0\,.
\nonumber
\eea
All hat objects are defined in 4 dimensions, whereas the
ones with tildes are $(n-4)$-dimensional remnants. 
Remaining integrals that contain remnants of the $n$-dimensional algebra, 
i.e. factors of $(\tilde k \cdot \tilde k)^\alpha$ are evaluated with the
following relations \cite{Binoth:2006hk}:
\bea
\int \frac{d^nk}{i \pi^{n/2}} \frac{(\tilde k \cdot \tilde k)^\alpha}{(k^2-M^2)^N} &=& 
 (-1)^{\alpha}  \frac{\Gamma(\alpha -\epsilon)}{\Gamma(1 -\epsilon)}
 \frac{n-4}{2}\; 
I_N^{n+2\alpha}\ ,\nonumber\\ 
\int \frac{d^nk}{i \pi^{n/2}} \frac{(\tilde k \cdot \tilde k)^\alpha k^\mu k^\nu}{(k^2-M^2)^N} &=&
(-1)^{\alpha+1} \frac{\Gamma(\alpha -\epsilon)}{\Gamma(1 -\epsilon)}
g^{\mu\nu} \frac{n-4}{4} 
\frac{n+2\alpha}{n} \,I_{N}^{n+2+2\alpha} \,.
\nonumber
\eea
The $n$-dimensional $N$-point integral is defined by \cite{Binoth:1999sp}
\bea
I_N^n &=& \int \frac{d^nk}{i \pi^{n/2}} \frac{1}{\prod_{j=1}^N((k-r_j)^2 - m_j^2)} \nonumber\\
      &=& (-1)^N \Gamma(N-n/2) \int\limits_0^1 dx_1 \dots dx_N
      \delta(1 - \sum\limits_{l=1}^N x_l ) 
( M^2 )^{n/2-N}\,, 
\nonumber
\eea
with $r_j=p_1+\dots +p_j$ and
\bea
 M^2 &=& \frac{1}{2}  \sum\limits_{i,j=1}^N x_i S_{ij} x_j \; , \;
 S_{ij} 
= -(r_i-r_j)^2 + m_i^2 + m_j^2 \,.
\nonumber
\eea
Below we list all required integrals, expanded to the
relevant order in $\epsilon$.
\bea
I_{4,\alpha =1}^{n,\mu\nu} &=&  \int \frac{d^nk}{i \pi^{n/2}}
\frac{\tilde k \cdot \tilde k \, 
k^\mu k^\nu}{\prod_{j=1}^4((k-r_j)^2 - m_j^2)} = -\frac{g^{\mu\nu}}{8} + {\cal O}(\epsilon)\nonumber\\
I_{4,\alpha=2}^{n} &=&  \int \frac{d^nk}{i \pi^{n/2}} 
\frac{(\tilde k \cdot \tilde k)^2}{\prod_{j=1}^4((k-r_j)^2 - m_j^2)}  
                     = -\frac{1}{6} +  {\cal O}(\epsilon)\nonumber\\
I_4^{n,\alpha=1} &=&  {\cal O}(\epsilon)\nonumber\\
I_{3,\alpha=1}^{n} &=&  \int \frac{d^nk}{i \pi^{n/2}} 
\frac{\tilde k \cdot \tilde k}{\prod_{j=1}^3((k-r_j)^2 - m_j^2)} 
 = -\frac{1}{2} + {\cal O}(\epsilon)\,
\nonumber
\eea

\end{appendix}


\newpage
\begin{thebibliography}{99}


\bibitem{Haywood:1999qg} 
  S.~Haywood {\it et al.}, ``Electroweak physics,''
  arXiv:hep-ph/0003275, published in the proceedings of the ``CERN
  Workshop on Standard Model Physics (and more) at the LHC'', 14-15
  October 1999, Geneva, Switzerland. Editors G.~Altarelli and
  M.L.~Mangano, Geneva, CERN, 2000.

\bibitem{Ohnemus:1991kk}  
  J.~Ohnemus,
  Phys.\ Rev.\ D {\bf 44} (1991) 1403.

\bibitem{Frixione:1993yp}
  S.~Frixione,
  Nucl.\ Phys.\ B {\bf 410} (1993) 280.

\bibitem{Ohnemus:1994ff}
  J.~Ohnemus,
  Phys.\ Rev.\ D {\bf 50} (1994) 1931 [arXiv:hep-ph/9403331].

\bibitem{Dixon:1998py}
  L.~J.~Dixon, Z.~Kunszt and A.~Signer,
  Nucl.\ Phys.\ B {\bf 531} (1998) 3 [arXiv:hep-ph/9803250].

\bibitem{Dixon:1999di}  
  L.~J.~Dixon, Z.~Kunszt and A.~Signer,
  Phys.\ Rev.\ D {\bf 60} (1999) 114037 [arXiv:hep-ph/9907305].

\bibitem{Campbell:1999ah}
  J.~M.~Campbell and R.~K.~Ellis,
  Phys.\ Rev.\ D {\bf 60} (1999) 113006 [arXiv:hep-ph/9905386].

\bibitem{Frixione:2006he}
  S.~Frixione and B.~R.~Webber,
  arXiv:hep-ph/0601192.

\bibitem{Grazzini:2005vw}
  M.~Grazzini,
  JHEP {\bf 0601}, 095 (2006)
  [arXiv:hep-ph/0510337].

\bibitem{Accomando:2004de}
  E.~Accomando, A.~Denner and A.~Kaiser,
  Nucl.\ Phys.\ B {\bf 706} (2005) 325
  [arXiv:hep-ph/0409247].

\bibitem{Dittmar:1996ss}
  M.~Dittmar and H.~K.~Dreiner,
  Phys.\ Rev.\ D {\bf 55} (1997) 167 [arXiv:hep-ph/9608317].

\bibitem{Dittmar:1996sp}
  M.~Dittmar and H.~K.~Dreiner,
  ``$h_0 \to W^+ W^- \to l^+ l'^- \nu/l \bar\nu/l'$ as the dominant SM Higgs
  search mode at the LHC for M($h_0$) = 155~GeV to 180~GeV,''
  arXiv:hep-ph/9703401, published in the proceedings of the Ringberg
  Workshop ``The Higgs Puzzle - What can We Learn from LEP2, LHC, NLC,
  and FMC?'', 8-13 December 1996, Ringberg, Germany. Editor B.A.
  Kniehl, Singapore, World Scientific, 1997.

\bibitem{Glover:1988fe}
  E.~W.~N.~Glover and J.~J.~van der Bij,
  Phys.\ Lett.\ B {\bf 219} (1989) 488.

\bibitem{Kao:1990tt}
  C.~Kao and D.~A.~Dicus,
  Phys.\ Rev.\ D {\bf 43} (1991) 1555.

\bibitem{Duhrssen:2005bz}
  M.~D\"uhrssen, K.~Jakobs, J.~J.~van der Bij and P.~Marquard,
  JHEP {\bf 0505} (2005) 064
  [arXiv:hep-ph/0504006].


\bibitem{Adamson:2002jb}
  K.~L.~Adamson, D.~de Florian and A.~Signer,
  Phys.\ Rev.\ D {\bf 65} (2002) 094041 [arXiv:hep-ph/0202132].

\bibitem{Adamson:2002rm}
  K.~L.~Adamson, D.~de Florian and A.~Signer,
  Phys.\ Rev.\ D {\bf 67} (2003) 034016 [arXiv:hep-ph/0211295].

\bibitem{Matsuura:1991pj}
  T.~Matsuura and J.~J.~van der Bij,
  Z.\ Phys.\ C {\bf 51} (1991) 259.

\bibitem{Zecher:1994kb}
  C.~Zecher, T.~Matsuura and J.~J.~van der Bij,
  Z.\ Phys.\ C {\bf 64} (1994) 219
  [arXiv:hep-ph/9404295].

\bibitem{Binoth:2005ua}
  T.~Binoth, M.~Ciccolini, N.~Kauer and M.~Kr\"amer,
  JHEP {\bf 0503} (2005) 065
  [arXiv:hep-ph/0503094].

\bibitem{BCKK_LH}
  T.~Binoth, M.~Ciccolini, N.~Kauer and M.~Kr\"amer, in 
  ``Les Houches physics at TeV colliders 2005, standard model, QCD, EW, and
  Higgs working group: Summary report,''
  arXiv:hep-ph/0604120.
  

\bibitem{xuetal} Z.~Xu, D.~Zhang, L.~Chang, Nucl. Phys. B291 (1987) 392.

\bibitem{'tHooft:1978xw}
  G.~'t Hooft and M.~J.~G.~Veltman,
  Nucl.\ Phys.\ B {\bf 153} (1979) 365.

\bibitem{Passarino:1978jh}
  G.~Passarino and M.~J.~G.~Veltman,
  Nucl.\ Phys.\ B {\bf 160} (1979) 151.

\bibitem{Binoth:2005ff}
  T.~Binoth, J.~P.~Guillet, G.~Heinrich, E.~Pilon and C.~Schubert,
  JHEP {\bf 0510} (2005) 015
  [arXiv:hep-ph/0504267].

\bibitem{Binoth:2006rc}
  T.~Binoth, M.~Ciccolini and G.~Heinrich,
  arXiv:hep-ph/0601254.
  
\bibitem{Binoth:1999sp}
  T.~Binoth, J.~P.~Guillet and G.~Heinrich,
  Nucl.\ Phys.\ B {\bf 572} (2000) 361 [arXiv:hep-ph/9911342].

\bibitem{Veltman:1988au}
  M.~J.~G.~Veltman,
  Nucl.\ Phys.\ B {\bf 319} (1989) 253.

\bibitem{Vermaseren:2000nd}  
  J.~A.~M.~Vermaseren,
  arXiv:math-ph/0010025 (unpublished).

\bibitem{Hahn:2000kx} 
  T.~Hahn,
  Comput.\ Phys.\ Commun.\ {\bf 140} (2001) 418 [hep-ph/0012260].

\bibitem{Hahn:1998yk}
  T.~Hahn and M.~Perez-Victoria,
  Comput.\ Phys.\ Commun.\  {\bf 118} (1999) 153
  [arXiv:hep-ph/9807565].

\bibitem{LesHouchesEventGeneratorInterface}
E.~Boos {\it et al.}, in proceedings of Workshop {\em Physics at TeV Colliders},
 Les Houches, France, 21 May - 1 June 2001, arXiv:hep-ph/0109068.

\bibitem{Berends:1994pv}
  F.~A.~Berends, R.~Pittau and R.~Kleiss,
  Nucl.\ Phys.\ B {\bf 424} (1994) 308
  [arXiv:hep-ph/9404313].

\bibitem{Kleiss:1994qy}
  R.~Kleiss and R.~Pittau,
  Comput.\ Phys.\ Commun.\  {\bf 83} (1994) 141
  [arXiv:hep-ph/9405257].

\bibitem{Kauer:2001sp}
  N.~Kauer and D.~Zeppenfeld,
  Phys.\ Rev.\ D {\bf 65} (2002) 014021
  [arXiv:hep-ph/0107181].

\bibitem{Kauer:2002sn}
  N.~Kauer,
  Phys.\ Rev.\ D {\bf 67} (2003) 054013
  [arXiv:hep-ph/0212091].

\bibitem{VEGAS}
G.~P. Lepage, J. Comput. Phys. {\bf 27} (1978) 192;
G.~P. Lepage, preprint CLNS-80/447, (1980).

\bibitem{OmniComp} \texttt{http://hepsource.sf.net/OmniComp/}

\bibitem{LHAPDF} \texttt{http://hepforge.cedar.ac.uk/lhapdf/}

\bibitem{GG2WW} \texttt{http://hepsource.sf.net/GG2WW/}

\bibitem{Buttar:2006zd}
  C.~Buttar {\it et al.},
   ``Les Houches physics at TeV colliders 2005, standard model, QCD, EW, and
   Higgs working group: Summary report,''
  arXiv:hep-ph/0604120.

\bibitem{Drollinger:2005ug}
  V.~Drollinger, T.~Binoth, M.~Ciccolini, M.~D\"uhrssen and N.~Kauer,
   ``Modeling the production of W pairs at the LHC,''
  CERN-CMS-NOTE-2005-024.

\bibitem{CMS-TDR}
  CMS Physics, Technical Design Report, CERN/LHCC 2006-021.

\bibitem{DittmarDreinerIII}
  M.~Dittmar and H.~K.~Dreiner, CMS-NOTE-1997-083 (unpublished).

\bibitem{JakobsTrefzger}
  K.~Jakobs, T.~Trefzger, ALTLAS-PHYS-2000-015 (unpublished).

\bibitem{Green:2000um}
  D.~Green, K.~Maeshima, J.~Marraffino, R.~Vidal, J.~Womersley, W.~Wu and S.~Kunori,
  J.\ Phys.\ G {\bf 26} (2000) 1751.
  
\bibitem{Davatz:2004zg} 
  G.~Davatz, G.~Dissertori, M.~Dittmar, M.~Grazzini and F.~Pauss,
  JHEP {\bf 0405} (2004) 009 [arXiv:hep-ph/0402218].

\bibitem{Djouadi:1997yw}
  A.~Djouadi, J.~Kalinowski and M.~Spira,
  Comput.\ Phys.\ Commun.\  {\bf 108} (1998) 56
  [arXiv:hep-ph/9704448].
 
\bibitem{Pumplin:2002vw} 
  J.~Pumplin, D.~R.~Stump, J.~Huston, H.~L.~Lai, P.~Nadolsky and
  W.~K.~Tung,
  JHEP {\bf 0207} (2002) 012 [arXiv:hep-ph/0201195].

\bibitem{Rainwater:1999sd}
  D.~L.~Rainwater and D.~Zeppenfeld,
  Phys.\ Rev.\ D {\bf 60} (1999) 113004
  [Erratum-ibid.\ D {\bf 61} (2000) 099901]
  [arXiv:hep-ph/9906218].

\bibitem{Dicus:1987fk}
  D.~A.~Dicus and S.~S.~D.~Willenbrock,
  Phys.\ Rev.\ D {\bf 37} (1988) 1801.

\bibitem{Dixon:2003yb}
  L.~J.~Dixon and M.~S.~Siu,
  Phys.\ Rev.\ Lett.\  {\bf 90} (2003) 252001
  [arXiv:hep-ph/0302233].

\bibitem{Catani:2001cr}
  S.~Catani, D.~de Florian and M.~Grazzini,
  JHEP {\bf 0201} (2002) 015
  [arXiv:hep-ph/0111164].

\bibitem{Binoth:2006hk}
  T.~Binoth, J.~P.~Guillet and G.~Heinrich,
  arXiv:hep-ph/0609054.

\end{thebibliography}
\end{document}